\documentstyle[12pt]{article}
\begin{document}
\textwidth=18cm
\textheight=22cm
\begin{center}{\large{\bf Nonequilibrium Phase Transition in the 
Kinetic \\
Ising model:Absence of tricritical behaviour \\
in presence of impurities}}\end{center}

\begin{center} {Muktish Acharyya}\end{center}

\begin{center} {Department of Physics, Presidency University,\\
86/1  College Street, Calcutta-700073, INDIA,\\
{\it muktish.physics@presiuniv.ac.in}}\end{center}

\vskip 2cm

The nonequilibrium dynamic phase transition,
in the two dimensional {\it site diluted} kinetic Ising model 
in presence of an oscillating magnetic field,
has been studied by Monte Carlo
simulation. 
The projections of dynamical phase boundary {\it surface} are drawn in the 
planes formed by the dilution and field amplitude and the plane formed
by temperature and field amplitude. The tricritical behaviour is found to be 
{\it absent} in this case which was observed in the pure system. 

\vskip 1cm

\noindent {\bf I. Introduction.}

Though the Ising model was proposed nearly three quarters of a century ago
its dynamical aspects are still under active investigation \cite{rev}. 
Nowadays,
the study of the dynamics of Ising models in presence of time varying magneic
field, became an active and interesting area of modern research.
The dynamical response of the Ising system in presence of an oscillating 
magnetic field has been studied
extensively by computer simulation 
\cite{rkp,dd,puri,tom,lo,ac} in the last few years. 
The dynamical hysteretic response 
\cite{rkp,dd,puri} and the nonequilibruim
dynamical phase transition \cite{tom,lo,ac} are two important aspects of
the dynamic response of the kinetic Ising model in presence of an
oscillating magnetic field.

Tome and Oliviera \cite{tom}first studied the dynamic transition in the
kinetic Ising model in presence of a sinusoidally oscillating magnetic field. 
They solved the mean field (MF)
dynamic equation of motion (for the average magnetisation) of the kinetic
Ising model in presence of a sinusoidally oscillating magnetic field.
By defining the order parameter as the time averaged
magnetisation over a full cycle of the 
oscillating magnetic field they showed that 
the order parameter vanishes
depending upon the value of the temperature and the 
amplitude of the oscillating field. 
Precisely, in the field amplitude and temperature plane they have drawn
a phase boundary separating dynamic ordered 
(nonzero value of order parameter) and disordered (order 
parameter vanishes) phase. They \cite{tom} have also observed 
and located a {\it tricritical point} (TCP),
(separating the nature (discontinuous/continuous) of the transition)
on the phase boundary line. It was confirmed later by Monte Carlo
study \cite{tcp} and by solving meanfield differential equation \cite{mftcp}
of kinetic Ising model. 

Since, this transition exists even in the static (zero frequency) limit
such a transition, observed \cite{tom}
 from the solution of mean field dynamical
equation, is not dynamic in true sense.
This is because, for the field amplitude less than the
coercive field (at temperature less than the transition temperature
without any field), the response magnetisation varies periodically
but asymmetrically even in the zero frequency limit; the system remains locked
to one well of the free energy and cannot go to the other one, in the absence
of noise or fluctuation.

Lo and Pelcovits \cite{lo} first attempted to study the dynamic 
nature of this phase transition (incorporating the effect of fluctuation)
in the kinetic Ising model by Monte Carlo (MC) simulation. 
In this case, the transition disappears in the 
zero frequency limit; due to the 
presence of fluctuations, the magnetisation flips to the
direction of the magnetic field and the dynamic order parameter (time
averaged magnetisation) vanishes.
However, they \cite{lo} have 
not reported any precise phase boundary.
Acharyya and Chakrabarti \cite{ac} studied the nonequilibrium dynamic phase
transition in the kinetic Ising model 
in presence of oscillating magnetic field by 
extensive MC simulation. 
It was also noticed by them \cite{ac}
that this dynamic phase transition is associated with the 
breaking of the symmetry of
the dynamic hysteresis ($m-h$) loop. 
In the dynamically disordered (value of order parameter vanishes)
phase the corresponding hysteresis loop is
symmetric, and loses its symmetry in the ordered phase (giving
nonzero value of dynamic order parameter).
They 
have \cite{ac}also studied the temperature variation of the ac susceptibility
components near the dynamic transition point.  
They observed 
that the imaginary (real) part of the ac susceptibility gives a
peak (dip) near the dynamic transition point (where the dynamic
order parameter vanishes). The conclusions were: (i) this is a distinct
signal of phase transition and (ii) this is an indication of the
thermodynamic nature of the phase transition.

It may be mentioned here that the statistical distribution of dynamic
order parameter has been studied by Sides et al \cite{rik}. The nature of the
distribution changes near the dynamic transition point. They have
also observed \cite{rik}
that the fluctuation of the hysteresis loop area becomes considerably
large near the dynamic transition point.

Very recently, 
the relaxation behaviour, of the dynamic order parameter,
near the transition point, has been studied
\cite{ma1}
by MC simulation and solving\cite{maab} meanfield 
dynamic equation. It has been observed 
that the relaxation is Debye type and the relaxation time diverges near the
transition point. The 'specific heat' and the 'susceptibility' also diverge
\cite{ma2}
near the transition point in a similar manner with that of fluctuations of
order parameter and fluctuation of energy respectively.

The tricritical point was observed in the case of pure system\cite{tcp}.
In this paper, the dynamic phase transition has been studied in the site
{\it diluted} (by nonmagnetic impurities)
kinetic Ising model by MC simulation. The phase boundaries
are plotted in the planes formed by the field amplitude and the temperature
and in the plane formed by the impurity 
concentration and the field amplitude. The paper
is organised as follows: in section II the model and the simulation scheme
are discussed, the MC results are given in section III and the paper ends
with a summary of the work in section IV.

\vskip 1cm

\noindent {\bf II. The Model and the simulation scheme}

The Hamiltonian, of a site diluted Ising model (with ferromagnetic nearest
neighbour interaction) in presence of a time varying magnetic field, can
be written as
\begin{equation}
H = -\sum_{<ij>} J_{ij} s_i^z s_j^z c_j c_j- h(t) \sum_i s_i^z c_i.
\label{hm} 
\end{equation}
Here, $s_i^z (=\pm 1)$ is Ising spin variable, $J_{ij}$ is the interaction
strength, $c_i$ (= 0 or 1) represents the site ($i$) which is either
occupied ($c_i$ = 1) or vacant ($c_i$ = 0).
$h(t) = h_0 {\rm cos}(\omega t)$  
is the oscillating magnetic field, where
$h_0$ and $\omega$ are the amplitude and the frequency 
respectively of the oscillating field. The system
is in contact with an isothermal heat bath at temperature $T$. For simplicity
all $J_{ij}$ and the value of the Boltzmann constant are taken equal to 
unity. The boundary condition is periodic.

A square lattice of linear size $L (=100)$ has been considered.
The lattice sites are randomly occupied by magnetic sites with a finite
probability $p$. So, the degree of dilution or the
concentration of (nonmagnetic) impurities,
is $q = 1-p$.
At any finite
temperature $T$ and for a fixed frequency ($\omega$) 
and amplitude ($h_0$) of the
field, the 
dynamics of this system has been studied here by Monte Carlo
simulation using Metropolis single spin-flip dynamics.
Each lattice site is
updated here sequentially and one such full scan over the lattice is
defined as the time unit (Monte Carlo step per spin or MCS)
here. The initial configuration has been chosen
such that the all spins are directed upward. The instanteneous magnetisation
(per site),
 $m(t) = (1/L^2) \sum_i s_i^z c_i$ has been calculated. From the instanteneous
magnetisation, the dynamic order parameter $Q = {\omega \over {2\pi}}
\oint m(t) dt$ (time averaged magnetisation over a full cycle of the
oscillating field) is calculated. This dynamic order parameter is a function of
temperature ($T$), field amplitude ($h_0$) and the impurity concentration
 ($q$) 
,i.e, $Q = Q(T,h_0,q)$. Each value of $Q$ has been calculated by averaging
over 25 number of initial impurity realisations. The frequency of the 
oscillating magnetic field used here is equal to 0.0628.

\vskip 1cm

\noindent {\bf III. Results}

It has been observed that $Q = Q(T,h_0,q)$ is nonzero for a finite set of
values of $h_0, T$ and $q$, and $Q$ vanishes elsewhere. In the space formed
by $h_0, T$ and $q$, there is a surface which divides the $Q = 0$ region from
$Q \neq 0$ region. Fig. 1 shows the schematic diagram of this phase surface.

Previously, a number of numerical studies [4,6,8] are 
performed regarding the dynamic
transition in the $q = 0$ plane (i.e., projection of this phase surface
on $h_0 -T$ plane. In that case, it was observed that in the $h_0-T$ plane
there is a distinct phase boundary below which $Q$ is nonzero and above which
$Q$ vanishes. There is a {\it tricritical point} on the phase boundary which 
separates the nature (discontinuous/continuous) of this transition.

Fig. 2a shows the dynamic phase boundary in the $h_0-T$ plane for different
values of the impurity concentration. 
It has been observed that as the impurity concentration increases the
phase boundary shrinks inward.
In this case, the entire phase boundary
has been scanned and the transition observed is always {\it continuous}. 
Unlike the earlier case \cite{ac}, no such {\it tricritical point} is
observed here.
Fig. 2b
shows the temperature variations 
of the dynamic order parameter (i.e, $Q$ versus $T$) for two different
values of $h_0$ in the case of very weak disorder (inpurity).

A similar kind of dynamic phase boundary has been obtained in the 
$h_0-q$ plane (i.e., the projection of the phase surface on $h_0-q$ plane).
Fig 3a. shows the phase boundaries in the $h_0-q$ plane for 
different values of temperatures. Like the
earlier case, here also, as the temperature increases the phase 
boundary shrinks inward. Here also the transition is continuous along the
entire phase boundary. Two typical transitions (fall of order parameter with
respect to the impurity concentrations) are shown in Fig. 3b.

\vskip 1cm

\noindent {\bf VI. Summary}

The nonequilibrium dynamic phase transition, in the 
{\it site diluted} kinetic Ising model
in presence of oscillating magnetic field, is studied by Monte
Carlo simulation.

The value of the dynamic order parameter gets nonzero below a boundary 
{\it surface} in $T$, $h_0$ and $q$ space, and above the surface it
vanishes. The projections of this surface on $h_0-T$ plane and
on $h_0-q$ plane are drawn. The nature of the transition observed here
is continuous along the entire phase boundary in the $h_0-T$ plane with
very small impurity concentrations. This is unlike the case observed 
earlier \cite{tcp,mftcp} in the pure sample. A similar kind of transition is 
observed for a fixed temperature with varying impurity concentrations.
Here also no tricritical behaviour was observed.
The studies reported here are mostly observational, no attempt has been
made to understand these phenomena from the 
knowledge of the theoretical background.

It should be mentioned here that a large scale simulation\cite{korniss}
observed that the first order transition is absent in the dynamic transition
in the pure Ising ferromagnet by oscillating magnetic field.

\newpage
\setlength{\unitlength}{0.240900pt}
\ifx\plotpoint\undefined\newsavebox{\plotpoint}\fi
\sbox{\plotpoint}{\rule[-0.200pt]{0.400pt}{0.400pt}}%


\bigskip

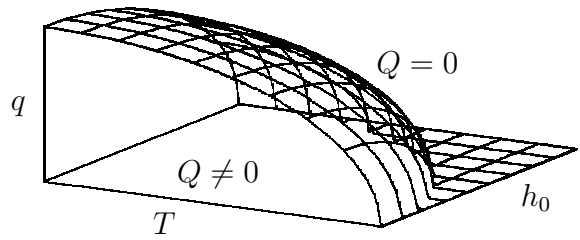
\noindent {\bf Fig. 1.} Schematic diagram of dynamic phase boundary surface in
the space formed by $T, h_0$ and $q$. Below this surface $Q \neq 0$ and above
the surface $Q = 0$.

\newpage
\setlength{\unitlength}{0.240900pt}
\ifx\plotpoint\undefined\newsavebox{\plotpoint}\fi
\sbox{\plotpoint}{\rule[-0.200pt]{0.400pt}{0.400pt}}%
%

\bigskip
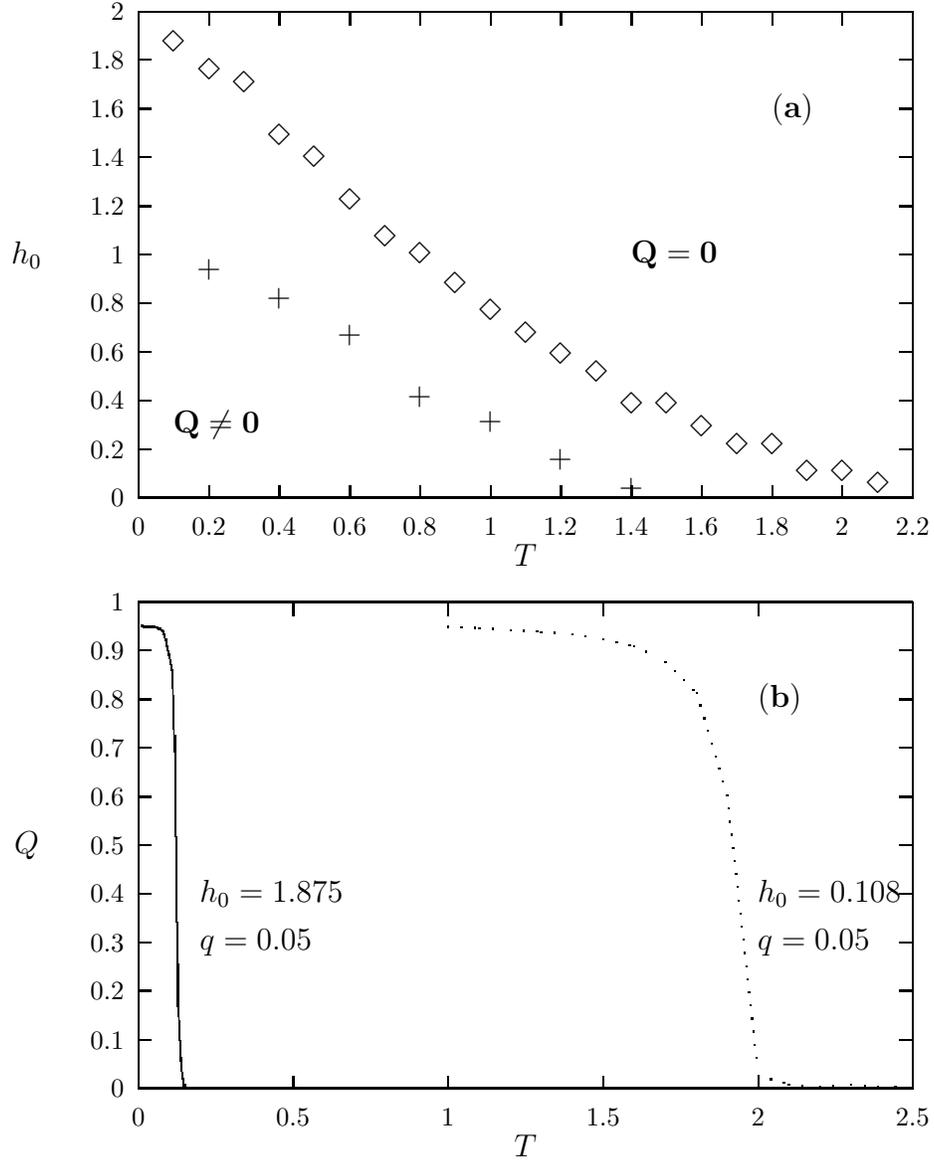
\noindent {\bf Fig. 2.} (a) Projections of dynamic phase surface on the $h_0-T$
plane. $(\Diamond)$ represents $q$ = 0.05 and $(+)$ represents $q$ = 0.3. (b)
Temperature variations of dynamic order parameter ($Q$) for two different 
values of field amplitudes ($h_0$).

\newpage
\setlength{\unitlength}{0.240900pt}
\ifx\plotpoint\undefined\newsavebox{\plotpoint}\fi
\sbox{\plotpoint}{\rule[-0.200pt]{0.400pt}{0.400pt}}%
%

\bigskip
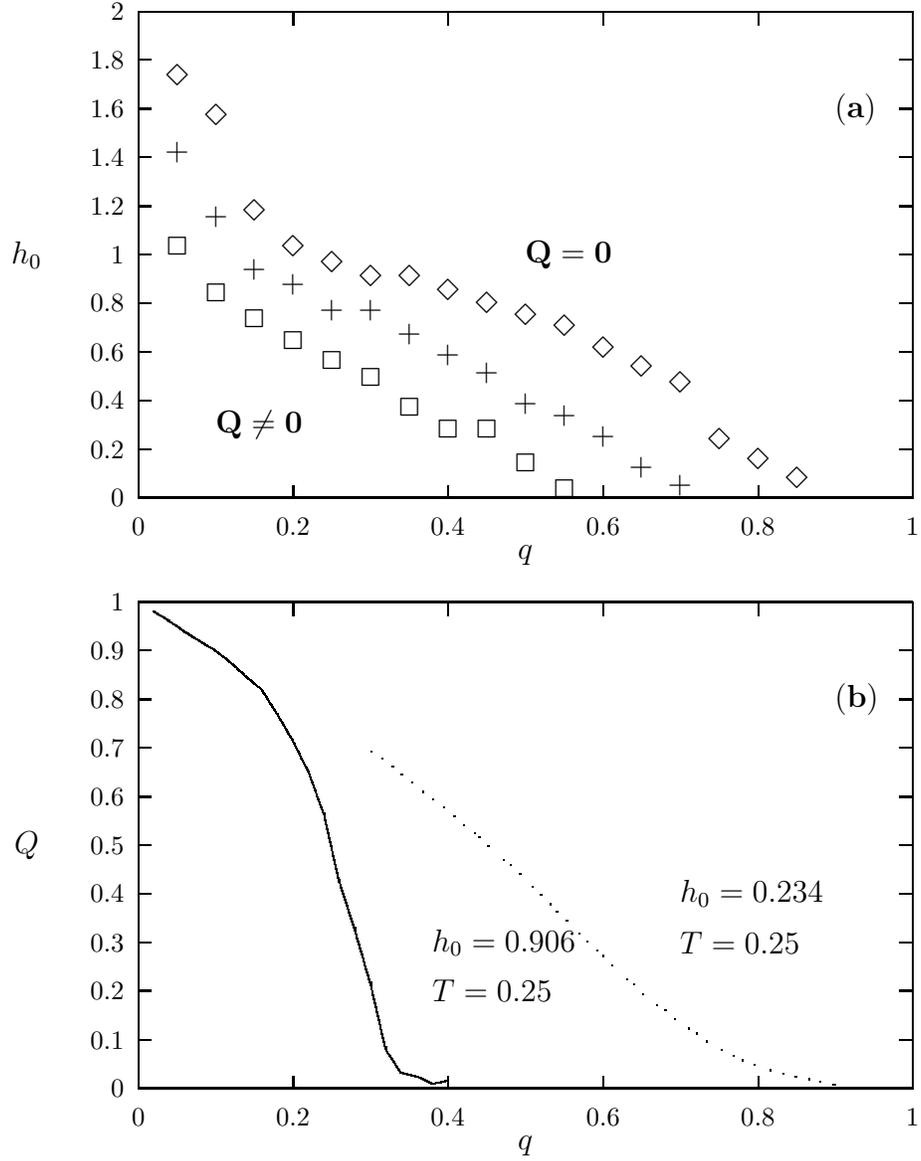
\noindent {\bf Fig. 3.} (a) Projections of dynamic phase surface on the $h_0-q$
plane. $(\Diamond)$ represents $T$ = 0.25, $(+)$ represents $T$ = 0.50
and ($\Box$) represents $T$ = 0.75. (b)
Variations of dynamic order parameter ($Q$) with respect to impurity
concentration ($q$) for two different 
values of field amplitudes ($h_0$).
\end{document}